\documentstyle[12pt]{article}
\input epsf.sty

\newcommand{\re}{{\rm e}}

\newcommand{\cO}{{\cal O}}

\newcommand{\cN}{{\cal N}}

\newcommand{\OO}{\mathop{\otimes}}
\font\cmss=cmss12 
\def\1{\hbox{{1}\kern-.25em\hbox{l}}}
\def\bfZ{\relax{\hbox{\cmss Z\kern-.4em Z}}}

\begin{document}
\begin{titlepage}

\centerline{\large \bf Exclusive evolution kernels in two-loop order:}
\centerline{\large \bf parity even sector.}

\vspace{20mm}

\centerline{\bf A.V. Belitsky\footnote{Alexander von Humboldt Fellow.},
                D. M\"uller}
\vspace{15mm}

\centerline{\it Institut f\"ur Theoretische Physik, Universit\"at
                Regensburg}
\centerline{\it D-93040 Regensburg, Germany}

\vspace{25mm}

\centerline{\bf Abstract}

\vspace{0.8cm}

We complete the construction of the non-forward evolution kernels in 
next-to-leading order responsible for the scale dependence of e.g.\ 
parity even singlet distribution amplitudes. Our formalism is designed 
to avoid any explicit two-loop calculations employing instead conformal 
and supersymmetric constraints as well as known splitting functions.

\vspace{6cm}

\noindent Keywords: evolution equation, two-loop exclusive kernels,
supersymmetric and conformal constraints

\vspace{0.5cm}

\noindent PACS numbers: 11.10.Hi, 11.30.Ly, 12.38.Bx

\end{titlepage}

\section{Introduction.}

Hard exclusive processes \cite{BroLep89}, i.e.\ involving a large
momentum transfer, provide a complimentary and an equally important
information about the internal structure of hadrons to the one gained
e.g.\ in deep inelastic scattering in terms of inclusive parton
densities. By means of QCD factorization theorems \cite{ColSopSte89}
physical observables measurable in these reactions, i.e.\ form factors
and cross sections, are expressed as convolution of a hard parton
rescattering subprocess and non-perturbative distribution amplitudes
\cite{BroLep89} and/or skewed parton distributions
\cite{MulRobGeyDitHor94,Ji96,Rad96,ColFraStr96}. Moreover, it is
implied that the main contribution to the latter comes from the
lowest two-particle Fock state in the hadron wave function. The
field-theoretical background for the study of the distribution
amplitudes is provided by their expression in terms of matrix elements
of non-local operators sandwiched between a hadron and vacuum states
(or hadron states with different momenta in the case of skewed parton
distributions)
\begin{equation}
\label{DefSPD}
\phi (x) = \frac{1}{2\pi} \int d z_- e^{i x z_-}
\langle 0 | \varphi^\dagger ( 0 ) \varphi ( z_- ) | h \rangle .
\end{equation}
Due to the light-like character of the path separating the partons,
$\varphi$, the operator in Eq.\ (\ref{DefSPD}) diverges in perturbation
theory and thus requires renormalization which inevitably introduces
a momentum scale into the game so that a distribution acquires a
logarithmic dependence on it. This dependence is governed by the
renormalization group which within the present context is cast into
the form of Efremov-Radyushkin-Brodsky-Lepage (ER-BL) evolution
equation \cite{EfrRad78,BroLep79}
\begin{equation}
\label{ER-BLequation}
\frac{d}{d \ln Q^2} \phi (x, Q) =
V \left(x, y | \alpha_s(Q) \right)
\OO^\re \phi (y, Q),
\qquad\mbox{with}\qquad
\OO^\re = \int_{0}^{1} dy .
\end{equation}
Note that the restoration of the generalized skewed kinematics in
perturbative evolution kernel, $V$, is unambiguous and straightforward
\cite{GeyDitHorMulRob88}. Therefore, we discuss in what follows only
the case when the skewedness of the process equals unity.

Recently, we have addressed the question of calculation of two-loop
approximation for the exclusive evolution kernels and give our results
for the parity-odd sector in Ref.\ \cite{BelMulFre99a}. The main tools
of our analysis were the constraints coming from known pattern of
conformal symmetry breaking in QCD and supersymmetric relations arisen
from super-Yang-Mills theory. In the present note we address the flavour
singlet parity even case which is responsible for the evolution of the
vector distribution amplitude.

\section{Anatomy of NLO evolution kernels.}

Our derivation is based on the fairly well established structure of the
ER-BL kernel in NLO. Up to two-loop order we have
\begin{equation}
\mbox{\boldmath$V$} (x, y | \alpha_s )
= \frac{\alpha_s}{2\pi}
\mbox{\boldmath$V$}^{(0)} (x, y)
+ \left( \frac{\alpha_s}{2\pi} \right)^2
\mbox{\boldmath$V$}^{(1)} (x, y)
+ \cO (\alpha_s^3) ,
\end{equation}
with the purely diagonal LO kernel $\mbox{\boldmath$V$}^{(0)}$ in the
basis of Gegenbauer polynomials and NLO one having the structure governed
by the conformal constraints \cite{Mue94,BelMul98a,BelMul98b}
\begin{equation}
\label{pred-Sing}
\mbox{\boldmath$V$}^{(1)}
= - \mbox{\boldmath$\dot V$} \OO^\re
\left(
\mbox{\boldmath$V$}^{(0)} + \frac{\beta_0}{2}\, \1
\right)
- \mbox{\boldmath$g$} \OO^\re \mbox{\boldmath$V$}^{(0)}
+ \mbox{\boldmath$V$}^{(0)} \OO^\re \mbox{\boldmath$g$}
+ \mbox{\boldmath$D$} + \mbox{\boldmath$G$}.
\end{equation}
Here the first three terms are induced by conformal symmetry breaking
counterterms in the $\overline{\mbox{MS}}$ scheme. Contrary to the LO
kernel $\mbox{\boldmath$V$}^{(0)} $, the so-called dotted kernel
$\mbox{\boldmath$\dot V$}^{(0)}$ and the $\mbox{\boldmath$g$}$ kernel
are off-diagonal in the space of Gegenbauer moments. They have been
obtained by a LO calculation \cite{Mue94,BelMul98a,BelMul98b}. The
remaining two pieces are diagonal and are decomposed into the
$\mbox{\boldmath$G$}^{V}$ kernel which is related to the crossed ladder
diagram and contains the most complicated structure in terms of Spence
functions, while the $\mbox{\boldmath$D$}^{V}$ kernel originates
from the remaining graphs and can be represented as convolution of
simple kernels.

One of the ingredients of the NLO result are the one-loop kernels.
We use for them improved expressions of Ref.\ \cite{BelMul98a} which
are completely diagonal in physical as well as unphysical spaces of
moments. In the matrix form we have
\begin{eqnarray}
\label{decomp-V-V}
\mbox{\boldmath$V$}^{(0)V} (x, y)
=
\left(
\begin{array}{ll}
C_F \left[ {^{QQ}\!v} (x, y) \right]_+
& - 2 T_F N_f \, {^{QG}\!v^V} (x, y) \\
C_F\, {^{GQ}\!v^V} (x, y)
& C_A {^{GG}\!v^V} (x, y)
- \frac{\beta_0}{2} \delta(x - y)
\end{array}
\right) ,
\end{eqnarray}
where $\beta_0 = \frac{4}{3} T_F N_f - \frac{11}{3} C_A$ and $C_A = 3$,
$C_F = 4/3$, $T_F = 1/2$ for QCD. The structure of the kernels reflects
the supersymmetry in $\cN = 1$ super-Yang-Mills theory
\cite{BukFroKurLip85,BelMulSch98,BelMul99a}
\begin{eqnarray}
\label{v-kernels}
&&{^{QQ}\!v} \equiv {^{QQ}\!v^a} + {^{QQ}\!v^b},
\quad
{^{QG}\!v^V} \equiv {^{QG}\!v^a} + 2 {^{QG}\!v^c},
\nonumber\\
&&{^{GQ}\!v^V} \equiv {^{GQ}\!v^a} + 2 {^{GQ}\!v^c} ,
\quad
{^{GG}\!v^V} \equiv \left[2\, {^{GG}\!v^a} + {^{GG}\!v^b} \right]_+
+ 2 {^{GG}\!v^c},
\end{eqnarray}
where the functions $v^i$ are defined in the following way
\begin{equation}
{^{AB} v^i}(x, y)
= \theta(y - x) {^{AB}\! f^i}(x, y)
\pm \left\{ {x \to \bar x \atop y \to \bar y } \right\}
\quad
\mbox{for}
\quad
\left\{ {A = B \atop A \not = B } \right. ,
\end{equation}
with (here and everywhere $\bar x\equiv 1 - x$)
\begin{eqnarray}
\left\{{ {^{AB}\! f^a} \atop {^{AB}\! f^b} }\right\}
&=& \frac{ x^{\nu(A) - 1/2}}{y^{\nu(B) - 1/2}}
\left\{ { 1 \atop \frac{1}{y - x} } \right\},
\nonumber\\
{^{AA}\! f^c}
&=& \frac{ x^{\nu(A) - 1/2}}{y^{\nu(A) - 1/2}}
\left\{
{ 2 \bar x y \left[ \frac{4}{3} - \ln( \bar x y ) \right] + y - x
\atop
2 \bar x y + y - x }
\right\}
\quad \mbox{for} \quad
A = \left\{ {Q \atop G } \right. ,
\nonumber\\
{^{AB}\! f^c}
&=& \frac{ x^{\nu(A) - 1/2}}{y^{\nu(B) - 1/2}}
\left\{
{ 2 x \bar y - \bar x
\atop 2 \bar x y - \bar y }
\right\}
\quad \mbox{for} \quad
A = \left\{ {Q \atop G} \right\} \not = B .
\end{eqnarray}
The index $\nu(A)$ coincides with the index of the Gegenbauer polynomials
in the corresponding channel, i.e.\ $\nu(Q) = 3/2$ and $\nu(G) = 5/2$.
The eigenvalues of the same $v^i$-kernel in different channels
are related to each other (here $v_{jj} \equiv v_j$)
\begin{eqnarray}
\label{eigenvalues-LO-a}
&&{^{QQ}\!v^a_j} = - \frac{1}{6} {^{QG}\!v^a_j} =
\frac{6}{j ( j + 3 )} {^{GQ}\!v^a_j} = \frac{1}{2} {^{GG}\!v^a_j}
= \frac{1}{(j + 1)(j + 2)},
\nonumber\\
\label{eigenvalues-LO-b}
&&{^{QQ}\!v^b_j} = {^{GG}\!v^b_j} - 1
= - 2 \psi( j + 2 ) + 2 \psi( 1 ) + 2,
\nonumber\\
\label{eigenvalues-LO-c}
&&{^{QQ}\!v^c_j} = - \frac{1}{6}{^{QG}\!v^c_j}
= \frac{6}{j ( j + 3 )}{^{GQ}\!v^c_j}
= \frac{1}{3}{^{GG}\!v^c_j}
=\frac{2}{j ( j + 1 )( j + 2 )( j + 3 )}.
\end{eqnarray}
Note that we have the identity
${^{GQ}\!v^c_j}={^{QQ}\!v^a_j}/3 = {^{GG}\!v^a_j}/6$, which in the
next section will serve as a guideline for the construction of
$\mbox{\boldmath$\dot V$}$ and $\mbox{\boldmath$G$}$ kernels.

\section{Construction of $\mbox{\boldmath$\dot V$}$ and
$\mbox{\boldmath$G$}$ kernels.}

To proceed further let us consider first the construction of the
so-called dotted kernels whose off-diagonal conformal moments are
simply expressed in terms of the one-loop anomalous dimensions,
${^{AB}\!\gamma}_j^{(0)}$, of the conformal operators as
$\theta_{j - 2,k} ({^{AB}\!\gamma}_j^{(0)} - {^{AB}\!\gamma}_k^{(0)})
d_{jk}$ with $d_{jk} = - \frac{1}{2}[ 1 + ( - 1)^{j - k} ]
\frac{(2k + 3)}{(j - k)(j + k + 3)}$. We introduce the matrix
\begin{equation}
\mbox{\boldmath$\dot V$}^{(0)V} (x, y)
=
\left(
\begin{array}{ll}
C_F \left[ {^{QQ}\!\dot v} (x, y) \right]_+
& - 2 T_F N_f {^{QG}\!\dot v}^V (x, y) \\
C_F {^{GQ}\!\dot v}^V (x, y)
&
C_A {^{GG}\!\dot v}^V (x, y)
\end{array}
\right) ,
\end{equation}
where we use the decomposition analogous to Eqs.\
(\ref{decomp-V-V},\ref{v-kernels})
for the LO kernels including the same ``+''-prescription although
this time the kernels are regular at the point $x = y$. The general
structure of ${^{AB} \dot v^i}$ reads
\begin{equation}
\label{DotKernel}
{^{AB} \dot v^i} (x, y) =
\theta(y - x) {^{AB}\! f^i} (x, y) \ln \frac{x}{y}
+ \Delta{^{AB}\! \dot{f}^i} (x, y)
\pm \left\{ {x \to \bar x \atop y \to \bar y } \right\} ,
\quad
\mbox{for}
\quad
\left\{ { A = B \atop A \not= B } \right. .
\end{equation}
For the dotted $a$ and $b$-kernels we have $\Delta{^{AB}\! \dot{f}^i}
(x, y) \equiv 0$ with $i = a, b$. To find the dotted $c$-kernels we make
use of the fact that kernels with the same conformal moments in different
channels are related by differential equations owing to the following
simple relations for the Gegenbauer polynomials
\begin{eqnarray}
\frac{d}{dx} C_j^{3/2}(2 x - 1)
= 6 C_{j-1}^{5/2}(2 x - 1),
\quad
\frac{d}{dx}  \frac{w(x|5/2)}{N_j(5/2)} C_{j - 1}^{5/2}(2 x - 1)
= - 6 \frac{w(x | 3/2)}{N_j (3/2)} C_{j}^{3/2}(2 x - 1),
\end{eqnarray}
were $w(x|\nu)=(x\bar{x})^{\nu - 1/2}$ is the weight function and
$N_j(\nu)= 2^{ - 4 \nu + 1 } \frac{ \Gamma^2 (\frac{1}{2}) \Gamma
( 2 \nu + j )}{\Gamma^2 (\nu) ( \nu + j ) j! }$ is the normalization
coefficient. From the knowledge of conformal moments, which
are determined by the eigenvalues of the corresponding kernels given
in Eq.\ (\ref{eigenvalues-LO-c}), and using the expansion of the
kernels w.r.t.\ the Gegenbauer polynomials
\begin{eqnarray*}
{^{AB}\!v^i} (x, y)
= \sum_{j = 0}^\infty
\frac{w \left( x | \nu(A) \right)}{N_j \left( \nu (A) \right)}
C^{\nu(A)}_{j + 3/2 - \nu(A)} (2 x - 1) {^{AB}\!v^i_j} \,
C^{\nu(B)}_{j + 3/2 - \nu(B)} (2 y - 1)
\end{eqnarray*}
we find then the following differential equations
\begin{eqnarray}
\frac{d}{dy} {^{GQ}\! \dot v^c} (x, y)
&=& {^{GG}\! \dot v^a} (x, y) +  {^{GG}\! v^a} (x,y),
\label{DiffEq1}\\
\frac{d}{dx} {^{GQ}\! \dot v^c} (x, y)
&=& - 2 \,{^{QQ}\! \dot v^a} (x, y) -{^{QQ}\! v^a} (x,y),
\label{DiffEq2}\\
{^{QG}\! \dot v^c} (x, y)
&=& \frac{1}{3} \frac{d}{dx} {^{GG}\! \dot v^c} (x, y) .
\label{DiffEq3}
\end{eqnarray}
One of the entries in Eq.\ (\ref{DotKernel}), namely
\begin{equation}
\Delta{^{GG}\! \dot{f}^c} (x, y) = 2 \frac{x^2}{y^2} (y - x),
\end{equation}
has been obtained in Ref.\ \cite{BelMul98b}. Thus defined
${^{GG}\!\dot v^c}$ kernel possesses the correct conformal moments
in both un- and physical sectors. Eqs.\ (\ref{DiffEq1},\ref{DiffEq2})
allow us (after fixing the integration constant) to find
$\Delta{^{GQ}\! \dot{f}^c}$, while from Eq.\ (\ref{DiffEq3}) we
conclude that $\Delta{^{QG}\! \dot{f}^c}$ is trivially induced by
$\Delta{^{GG}\! \dot{f}^c}$. Therefore, we have finally
\begin{eqnarray}
\Delta{^{GQ}\! \dot{f}^c}
= x^2 (2 x - 3) \ln\frac{x}{y},
\qquad
\Delta{^{QG}\! \dot{f}^c}
= - \frac{x}{3 y^2} \left( 4 x - 5  y + 2 x y \right).
\end{eqnarray}

Next the $\mbox{\boldmath$g$}$ function is given by
\cite{BelMul98a,BelMul98b}
\begin{eqnarray}
\label{set-g-kernels}
\mbox{\boldmath$g$} (x, y) =
\theta(y - x)
\left(
\begin{array}{cc}
- C_F \left[ \frac{ \ln \left( 1 - \frac{x}{y} \right) }{y - x} \right]_+
& 0 \\
C_F \frac{x}{y}
& - C_A\left[ \frac{ \ln \left( 1 - \frac{x}{y} \right) }{y - x} \right]_+
\end{array}
\right)
\pm
\left\{ x \to \bar x \atop y \to \bar y \right\},
\end{eqnarray}
with ($-$) $+$ sign corresponding to (non-) diagonal elements.

The construction of the diagonal $\mbox{\boldmath$G$} (x, y)$ kernel
related to the crossed ladder diagrams is straightforward up to a
number of points which are not obvious and present the most non-trivial
part of the machinery. Let us give here its construction in more
detail as compared to Ref.\ \cite{BelMulFre99a}. From the result
in the flavour non-singlet sector \cite{Sar84,DitRad84,MikRad85}
and the general limiting procedure to the forward case
\cite{MulRobGeyDitHor94,GeyDitHorMulRob88} we know that all entries
in the matrix
\begin{equation}
\label{G-kernel-odd}
\mbox{\boldmath$G$}^i (x, y)
= - \frac{1}{2}
\left(
\begin{array}{cc}
2 C_F \left( C_F - \frac{C_A}{2} \right)
\left[ {^{QQ}\!G}^i (x, y) \right]_+
&
2 C_A T_F N_f \, {^{QG}\!G}^i (x, y)
\\
C_F C_A \, {^{GQ}\!G}^i (x, y)
&
C_A^2 \left[ {^{GG}\!G}^i (x, y) \right]_+
\end{array}
\right) ,
\end{equation}
must have the following general structure
\begin{equation}
\label{kernel-G}
{^{AB} G}^i (x, y)
= \theta (y - x)
\left( {^{AB}\! H}^i + \Delta{^{AB}\! H}^i \right) (x, y)
+ \theta (y - \bar x)
\left( {^{AB} \overline H}^i
+ \Delta{^{AB} \overline H}^i \right) (x, y),
\end{equation}
with the following expressions for $H$ and $\overline{H}$
\begin{eqnarray}
\label{kernel-S-H}
{^{AB} H}^i (x, y)
\!\!&=&\!\! 2 \left[ \pm {^{AB} \overline f}^i
\left( {\rm Li}_2( \bar x ) + \ln y \ln \bar x \right)
- {^{AB}\! f}^A\, {\rm Li}_2( \bar y ) \right],
\\
\label{kernel-S-bH}
{^{AB} \overline{H}}^i (x, y)
\!\!&=&\!\! 2 \left[
\left( {^{AB}\! f}^i \mp {^{AB} \overline f}^i \right)
\left( {\rm Li}_2 \left( 1 - \frac{x}{y} \right)
+ \frac{1}{2} \ln^2 y \right)
+ {^{AB}\! f}^i \left( {\rm Li}_2 ( \bar y )
- {\rm Li}_2 (x) - \ln y \ln x \right) \right],
\nonumber\\
\end{eqnarray}
where the upper (lower) sign corresponds to the $A = B$ ($A \not= B$)
channels. For the $QQ$ sector we have $\Delta{^{QQ}\!H} =
\Delta{^{QQ}\!\overline H}=0$. However, in general these addenda
are nonzero and are needed to ensure the diagonality of the kernels.
From the known two-loop splitting functions we have to require as well
that in the forward limit these terms contribute only to rational
functions and/or terms containing single logs of momentum fractions.

The reduction $P \to V^{\rm D}$ procedure from the forward to non-forward
kinematics \cite{BelMul98a} is hard to handle for the restoration of
$\Delta H$ contributions, so we have to rely on different principles.
We do this by exploring supersymmetry and conformal covariance of $\cN = 1$
super-Yang-Mills theory \cite{BelMulSch98,BelMul99a}. As a matter of fact
being wrong for all order results these assumptions hold true within the
present context since the ${^{AB} G}$ kernels arise from the crossed
ladder diagrams which have no UV divergent subgraphs and, therefore,
require no renormalization. Thus, these kernels can be constructed from
six constraints on anomalous dimensions of conformal operators. In
principle these relations can also be written for the kernels in the ER-BL
representation so that taking the known two entries of the quark-quark
channel one can deduce all other channels. Unfortunately, at first glance
it seems that not all of these constraints have a simple solution in the
ER-BL representation. For this reason we modify our construction in the
following way. Because of both supersymmetry and conformal covariance, the
mixed channels are related by the equation
\begin{eqnarray}
\label{const-III}
{^{GQ}\!G^i} (x, y)
= \frac{(\bar x x)^2}{\bar y y} {^{QG}\!G^i} (y, x).
\end{eqnarray}
Employing this relation we can get a further one, from the so-called
Dokshitzer supersymmetry constraint,
\begin{eqnarray}
\label{const-I+III}
\frac{d}{dy}
{^{QQ}\!G^i} (x, y)
+ \frac{d}{dx} {^{GG}\!G^i} (x, y)
= - 3 {^{QG}\!G^i} (x, y),
\end{eqnarray}
which allows to obtain the $GG$ entry provided we already know the
kernel in the mixed channel.

Let us consider first the parity odd sector. At LO we have for moments
${^{QG}v^A_j} = 6 {^{QQ}\!v^a_j}$. Thus we can simply obtain the $QG$
kernel differentiating the $QQ$ one. Fortunately, it turns out that
the ${^{QG}\!G^A}$ kernel can be obtained\footnote{The correctness of
this and subsequent results is checked by forming the Gegenbauer
moments and comparing them with known NLO forward anomalous dimensions
\cite{FurPet80}.} in the same way
\begin{eqnarray}
{^{QG}\!G^A} (x, y) = \frac{d}{dy} {^{QQ}\!G^a} (x, y),
\end{eqnarray}
where ${^{QQ}\!G^a}$ is given by Eqs.\ (\ref{kernel-G}-\ref{kernel-S-bH})
with $\Delta{^{QQ}\!H}^a = \Delta{^{QQ}\!\overline H}^a = 0$. The $GQ$
entry is simply deduced from Eq.\ (\ref{const-III}), while the $GG$ one
comes from the solution of the differential equation (\ref{const-I+III}).
The integration constant as a function of $y$ is almost fixed by the
necessary condition of diagonality
\begin{eqnarray}
{^{GG}\!G^A}(x,y)
= \frac{(x \bar x)^2}{(y \bar y)^2} {^{GG}\!G^A}(y,x) .
\end{eqnarray}
The remaining degree of freedom can be easily fixed from the requirement
that the moments ${^{GG}\!G}^A_{ji}$ are diagonal for $i = 0, 1$. To
simplify the result, we remove a symmetric function (w.r.t.\ the
simultaneous interchange $x \to \bar x$ and $y \to \bar y$) which enters
in both $\Delta{^{GG}\!H}^A$ and $\Delta{^{GG}\!\overline H}^A$ kernels,
however with different overall signs and, therefore, disappears from
${^{GG}\!G}^A$. We present our final results in a symmetric manner as
(cf.\ \cite{BelMulFre99a})
\begin{eqnarray}
\Delta{^{QQ}\!H}^A (x, y)
&=& \Delta{^{QQ}\!\overline H}^A (x, y) = 0,
\\
\Delta {^{QG}\!H}^A (x, y)
&=& \Delta {^{QG}\!\overline H}^A (\bar x, y),
\quad
\Delta{^{QG}\!\overline H}^A (x, y)=
 \frac{x \bar x}{(y \bar y)^2} \Delta{^{GQ}\!\overline H}^A (y,x)
\\
\Delta{^{GQ}\!H}^A (x, y)
&=&  - \Delta{^{GQ}\!\overline H}^A (\bar x, y)
\quad
\Delta{^{GQ}\!\overline H}^A (x, y)
=
 - 2 \frac{x \bar x}{y} \ln x + 2 \frac{x \bar x}{\bar y} \ln y,
\\
\Delta{^{GG}\!H}^A (x, y)
&=& -\Delta{^{GG}\!\overline H}^A(\bar x, y),
\\
\Delta{^{GG}\!\overline H}^A (x, y)
&=&
\frac{1 - 2x \bar x}{4 \bar y^2} + \frac{1 - 2 \bar x (1 + \bar x)}{4 y^2}
- 2 \frac{x (\bar x + y - 3 \bar x y)}{\bar y y^2} \ln x
- 2 \frac{\bar x (x + \bar y - 3 x \bar y)}{y \bar y^2} \ln y .
\nonumber
\end{eqnarray}

Now instead of dealing with the whole parity even sector, we can
consider only the difference between vector and axial-vector functions
\begin{equation}
H^V = H^A + H^\delta .
\end{equation}
In LO we have the simple equation ${^{GQ}\!v^c_j} = {^{QQ}\!v^a_j}/3
= {^{GG}\!v^a_j}/6$, see Eq.\ (\ref{eigenvalues-LO-c}), which allows
us to write down a simple relation between the kernels in different
channels. However, to preserve the generic form of the ${^{GQ}\!G^c}$
function in the forward limit \cite{FurPet80} we have used the following
modified differential equations\footnote{Note that we introduce a
shorthand notation for the convolution, namely, ${^{QQ}\!f^i} \OO^\re
{^{QQ}\!f^j}$ is understood as convolution of the corresponding ER-BL
kernels and then keeping only the part proportional to $\theta(y-x)$.}:
\begin{eqnarray}
\frac{d}{dx} {^{GQ}\!H^c}
= - 4 \left( {^{QQ}\! H^a} + 9\, {^{QQ}\! f^c}
\OO^\re
{^{QQ}\! f^c} \right),
\quad
\frac{d}{dy} {^{GQ}\! H^c}
= 2 \left( {^{GG}\! H^a} + 2\, {^{GG}\! f^c}
\OO^\re
{^{GG}\! f^c} \right),
\nonumber\\
\frac{d}{dx} {^{GQ}\!\overline{H}^c}
= - 4 \left( {^{QQ}\!\overline{H}^a}
+ 9\, {^{QQ}\!\tilde{f}^c} \OO^\re {^{QQ}\! f^c} \right),
\quad
\frac{d}{dy} {^{GQ}\! \overline{H}^c}
= 2 \left( {^{GG}\! \overline{H}^a}
+ 2\, {^{GG}\! \tilde{f}^c} \OO^\re {^{GG}\! f^c} \right),
\end{eqnarray}
where $\tilde{f}^c (x, y) \equiv f^c (\bar x, y)$. The kernels
${^{GG}\! H^a}$ and ${^{GG}\! \overline H^a}$ are the parts of the
whole parity odd functions derived in the fashion already explained
above. The two sets of differential equations can be solved up to
two integration constants which can easily be fixed from the
diagonality of their conformal moments. Finally, we simplify the
solution by adding pure diagonal pieces containing $a$ and $c$ kernels
and their convolution as well as by removing symmetric terms which die
out in ${^{GQ}\!G}$.

The entry in the $QG$ channel can be obtained from the supersymmetric
relation (\ref{const-III}). To construct the $GG$ kernel we use then
the constraint (\ref{const-I+III}) with ${^{QQ}\! G^c} \equiv 0$.
We determine the integration constant as a function of $x$ in the
same manner as described previously. Our findings for the
$\Delta{^{AB} H}^\delta$ and $\Delta{^{AB}\overline H}^\delta$ can
be summarized in the formulae
\begin{eqnarray}
\Delta{^{QG}\! H}^\delta (x, y)
&=& - \frac{x\bar x}{(y \bar y)^2}
\Delta{^{GQ}\!H}^\delta (\bar y, \bar x),
\quad
\Delta{^{QG}\!\overline H}^\delta (x, y)
= \frac{x \bar x}{(y \bar y)^2}
\Delta{^{GQ}\!\overline H}^\delta (y, x),
\\
\Delta {^{GQ}\!H}^\delta (x, y)
&=& \Delta {^{GQ}\!\overline H}^\delta (\bar x, y)
+ 20 \frac{x (x - \bar x)}{3y}
- 4 \frac{\bar x (3 + 2 \bar x)}{3y} \ln\bar x
+ 4 \frac{x (3 + 2 x)}{3\bar y} \ln y,
\\
\Delta{^{GQ}\!\overline H}^\delta (x, y)
&=&- \frac{61}{18}
+ 2 x \bar x \Big( 1 - ( 3 - 10 \bar x ) \ln y
+ ( 3 - 10 x ) \ln x \Big)
\nonumber\\
&+& \frac{\bar x \left( 6 - 19 \bar x + 6 \bar x^2 \right)}{3y}
- 2\frac{\bar x \left( y + x ( \bar x - x) \right)}{\bar y} \ln y
+ 2 \frac{x \left( \bar y + \bar x ( x - \bar x) \right)}{y} \ln x ,
\nonumber\\
\Delta{^{GG}\!H}^\delta (x, y)
&=& \Delta{^{GG}\!\overline H}^\delta (\bar x, y)
- \frac{20 - 18 x + 55 x \bar x}{6 y^2}
- \frac{20 - 23 x \bar x}{6 \bar y^2}
- \frac{17 + 32 x + 28 x^2}{6 y \bar y} \\
&-& 2 \frac{\bar x}{y}
\left( 2 \frac{ \bar x - x }{\bar y}
+ \frac{2 + 3 \bar x}{y} \right) \ln\bar x
- 2 \frac{x}{\bar y}
\left(
2 \frac{x - \bar x}{y}
+ \frac{2 + 3 x}{\bar y} \right)\ln y,
\nonumber\\
\Delta{^{GG}\!\overline H}^\delta (x, y)
&=&
-(1 - x - y) \left(
\frac{20 - 22 x + 21 x \bar x}{6 y^2}
+ \frac{20 - 22 \bar x + 21 x \bar x}{6 \bar y^2}
+ \frac{39 + 38 x \bar x}{6 y \bar y}
\right)
\nonumber\\
&+& 2 \left( \frac{x^3}{3 y^2}
- \frac{x^2 (21 - 20 x) }{3 y}
-2 \frac{x \bar x^2}{\bar y} \right)\ln x
+ 2 \left(
\frac{\bar x^3}{3\bar y^2}
- \frac{\bar x^2 (21 - 20 \bar x) }{3\bar y}
- 2 \frac{\bar x x^2}{y}
\right)\ln y .
\nonumber
\end{eqnarray}

\section{Restoration of remaining diagonal terms.}

As in the twist-two axial and transversity sectors \cite{BelMulFre99a}
it turns out that the remaining diagonal piece, $\mbox{\boldmath$D$}^V$,
can be represented as the convolution of simple diagonal kernels. To
find it we take first the forward limit
\begin{eqnarray*}
\label{SingletLimit}
\mbox{\boldmath$P$} (z)
= {\rm LIM}\, \mbox{\boldmath$V$} (x, y)
\equiv \lim_{\tau\to 0} \frac{1}{|\tau|}
\left(
\begin{array}{rr}
{^{QQ} V}
&
\frac{1}{\tau}{^{QG} V}
\\
\frac{\tau}{z} {^{GQ} V}
&
\frac{1}{z}{^{GG} V}
\end{array}
\right)^{\rm ext}
\left( \frac{z}{\tau}, \frac{1}{\tau} \right) ,
\end{eqnarray*}
and compare our result with the known DGLAP kernel
$\mbox{\boldmath$P$}^V$ \cite{FurPet80}. In this way,
\begin{eqnarray}
\mbox{\boldmath$D$}^V (z)
= \mbox{\boldmath$P$}^V(z)
- {\rm LIM}
\left\{
- \mbox{\boldmath$\dot V$} \OO^\re
\left( \mbox{\boldmath$V$}^{(0)V} + \frac{\beta_0}{2} \1 \right)
- \mbox{\boldmath$g$} \OO^\re \mbox{\boldmath$V$}^{(0)V}
+ \mbox{\boldmath$V$}^{(0)V} \OO^\re \mbox{\boldmath$g$}
+ \mbox{\boldmath$G$}^{V}
\right\},
\end{eqnarray}
we extract the remaining part ${\rm LIM} \mbox{\boldmath$D$}^V(x,y)$
and find then the desired convolutions in the forward representation.
Note that we map the antiparticle contribution, i.e. $z < 0$, into the
region $z > 0$ by taking into account the underlying symmetry of the
singlet parton distributions. Our findings can be immediately mapped
back into the ER-BL representation:
\begin{eqnarray}
\label{D-QQ-e}
{^{QQ}\! D}^V
&=& C_F^2 \left[ D_F \right]_+
- C_F \frac{\beta_0}{2} \left[ D_\beta \right]_+
- C_F \left( C_F - \frac{C_A}{2} \right)
\left[ \frac{4}{3} {^{QQ}\!v} + 2\, {^{QQ}\!v}^a \right]_+ \\
&+& 4\, C_F T_F N_f \left\{
\frac{1}{3} {^{QQ}\!v}^a \OO^\re {^{QQ}\!v}^a
- 6{^{QQ}\!v}^c \OO^\re {^{QQ}\!v}^c
- {^{QQ}\!v}^a + \frac{7}{6}{^{QQ}\!v}^c
\right\}, \nonumber\\
{^{QG}\!D}^V
&=& - C_F T_F N_f
\left\{
2\left[ {^{QQ}\!v} \right]_+  \OO^\re {^{QG}\!v}^c
+ {^{QQ}\!v}^a \OO^\re {^{QG}\!v}^a
+ \frac{3}{2} {^{QG}\!v}^a + 6 {^{QG}\!v}^a
\right\} \\
&+& 2\, C_A T_F N_f
\Bigg\{
- \left[ \frac{8}{3} \left[{^{QQ}\!v} \right]_+
+ 56 {^{QQ}\!v}^c \right]
\OO^\re {^{QG}\!v}^c + \frac{130}{3} {^{QQ}\!v}^a
\OO^\re {^{QG}\!v}^a \nonumber\\
&&\hspace{2cm}+ \left[ \frac{55}{9} - 2 \zeta (2) \right] {^{QG}\!v}^a
- \left[\frac{301}{18} + 4 \zeta (2) \right] {^{QG}\!v}^a
\Bigg\}, \nonumber\\
{^{GQ}\!D}^V
&=& C_F^2
\left\{
-\left[ {^{GG}\!v}^A \right]_+
\OO^\re \left[\frac{1}{2} {^{GQ}\!v}^a
+ 3 {^{GQ}\!v}^c \right]
-5 {^{GG}\!v}^a  \OO^\re {^{GQ}\!v}^a
- 3 {^{GQ}\!v}^a
\right\} \nonumber\\
&-& C_F \beta_0
\left\{\left[ {^{GG}\!v}^A \right]_+  \OO^\re
\left[\frac{1}{2} {^{GQ}\!v}^a +  {^{GQ}\!v}^c \right]
+ \frac{3}{4} {^{GG}\!v}^a  \OO^\re {^{GQ}\!v}^a
+ \frac{5}{3} {^{GQ}\!v}^a
\right\} \\
&+& C_F C_A
\Bigg\{
- \left[{^{GG}\!v}^A \right]_+ \OO^\re
\left[ {^{GQ}\!v}^a - \frac{3}{2} {^{GQ}\!v}^c \right]
-\frac{25}{6} {^{GG}\!v}^a \OO^\re {^{GQ}\!v}^a
+ 9 {^{GG}\!v}^c \OO^\re {^{GQ}\!v}^c
\nonumber\\
&&\hspace{2cm} - \left( \frac{43}{9}
+ 2 \zeta(2) \right) {^{GQ}\!v}^a
+ \left( \frac{8}{9} - 4 \zeta(2) \right) {^{GQ}\!v}^c
\Bigg\} , \nonumber\\
{^{GG}\!D}^V
&=& C_A^2
\Bigg\{
\left[ {^{GG}\!v}^A \right]_+
\OO^\re \left[ {^{GG}\!v}^a + \frac{11}{3} {^{GG}\!v}^c \right]
- 14 {^{GG}\!v}^a \OO^\re {^{GG}\!v}^a
+ 12{^{GG}\!v}^c \OO^\re {^{GG}\!v}^c \nonumber\\
&+& \frac{2}{3} \left[ {^{GG}\!v}^A \right]_+
- \frac{131}{12} {^{GG}\!v}^a
+ \frac{91}{18} {^{GG}\!v}^c - 2 \delta(x - y)
\Bigg\} \\
&-& C_A \frac{\beta_0}{2}
\left\{
- \frac{1}{2} {^{GG}\!v}^a \OO^\re {^{GG}\!v}^a
+ \frac{5}{3} \left[ {^{GG}\!v}^V \right]_+
+ 3 {^{GG}\!v}^a + \frac{13}{3} {^{GG}\!v}^a + 2 \delta(x - y)
\right\} \nonumber\\
&+& C_F T_F N_f
\left\{
{^{GG}\!v}^a \OO^\re {^{GG}\!v}^a
+ \frac{4}{3} {^{GG}\!v}^c -\delta(x-y)
\right\} . \nonumber
\end{eqnarray}
where $D_F$, $D_\beta$ functions are known from the flavour non-singlet
case \cite{BelMulFre99a}. In comparison to the parity odd sector
the convolution of $c$-kernels appears as a new entry. It is worth
mentioning that our result for the evolution kernels in the parity even
singlet sector possesses the correct conformal moments in both the
physical and unphysical sectors. This is to be contrasted with an
explicit momentum fraction space calculation at LO and quark bubble
insertions in NLO kernels for the mixed channels \cite{BelMul98a}
where the improved kernels do not appear.

\section{Conclusions.}

To recapitulate the results presented here, we have reconstructed
the two-loop singlet evolution kernels responsible for the scale
dependence of the vector meson distribution amplitudes. We have
avoided cumbersome next-to-leading calculations by adhering to an
extensive use of the conformal and supersymmetric constraints derived
earlier which thus play a paramount r\^ole in the formalism. The
correctness of the results given here is proved by evaluating
the Gegenbauer moments of the kernels which coincide with the
anomalous dimensions derived in Ref.\ \cite{BelMul98b}. Our findings
allow to use now the direct numerical integration (see Ref.\
\cite{FFGS98,MusRad99} for a leading order analysis of non-forward
parton distributions) of the generalized exclusive evolution equations
which provides a competitive alternative to the previously developed
methods of orthogonal polynomial reconstruction of skewed parton
distribution pursued by us in Ref.\ \cite{Beletal97}.

\vspace{1cm}

A.B. was supported by the Alexander von Humboldt Foundation.

\end{document}